\documentclass[aps,pra,twocolumn]{revtex4}

\usepackage{graphicx,epsfig}
\usepackage{times}
\usepackage{amsmath,amssymb,wasysym}
\usepackage{color}

\usepackage[normalem]{ulem} 
\usepackage[utf8]{inputenc}
\usepackage[linktocpage=true,colorlinks=true, pdfborder={0 0 0},linkcolor=blue,citecolor=blue,urlcolor=blue,anchorcolor=black,linkcolor=blue]{hyperref}
\usepackage[twoside, bindingoffset=0mm, left=20mm, right=20mm, top=24mm, bottom=24mm,a4paper]{geometry} 

\newcommand{\quoting}[1]{``#1''}

\newcommand{\rem}[1]{}

\newcommand{\mean}[1]{\langle#1\rangle}
\newcommand{\test}{t}
\newcommand{\AQDs}{A_{\circ}}
\newcommand{\AWL}{A}
\newcommand{\emax}{e_{\text{max}}}
\newcommand{\numax}{\nu_{\text{max}}}
\newcommand{\Eco}{E_{\text{cut-off}}}
\newcommand{\dWL}{d_{\text{WL}}}

\begin{document}

\title{Morphology of wetting-layer states in a simple quantum-dot wetting-layer model}
  \author{Marcel Eichelmann}
  \affiliation{Institut f{\"u}r Physik, Otto-von-Guericke-Universit{\"a}t Magdeburg, Postfach 4120, D-39016 Magdeburg, Germany}
  \author{Jan Wiersig}
\email{jan.wiersig@ovgu.de}
  \affiliation{Institut f{\"u}r Physik, Otto-von-Guericke-Universit{\"a}t Magdeburg, Postfach 4120, D-39016 Magdeburg, Germany}
\date{\today}

\begin{abstract}
The excitation of semiconductor quantum dots often involves an attached wetting layer with delocalized single-particle energy eigenstates. These wetting-layer states are usually approximated by (orthogonalized) plane waves. 
We show that this approach is too crude. Even for a simple model based on the effective-mass approximation and containing one or a few lens-shaped quantum dots on a rectangular wetting layer, the wetting-layer states typically show a substantially irregular and complex morphology. To quantify this complexity we use concepts from the field of quantum chaos such as spectral analysis of energy levels, amplitude distributions, and localization of energy eigenstates.
\end{abstract}
\maketitle
\section{Introduction}
\label{sec:intro}
Self-assembled semiconductor quantum dots (QDs) have attracted a lot of attention due to their potential for fundamental research, such as superradiance~\cite{JGA16} and cavity-quantum electrodynamics~\cite{LFNIOVV04}, as well as device applications, such as single-photon sources~\cite{Michler2000c} and low-threshold microlasers~\cite{WGJ09}. 
QDs allow for a carrier confinement in all three dimensions on the nanometer scale with a discrete atomic-like density of states. The corresponding single-particle energy eigenstates are spatially strongly localized at the respective QD position. The morphology of such QD states has been studied comprehensively in the literature with pseudopotential theory~\cite{BNZ03,BZ03}, ${\bf k \cdot p}$ models~\cite{RSPGB05}, tight-binding models~\cite{SCH05,Norman2}, and density functional theory~\cite{NM00}. 

In self-assembled QDs grown in the Stranski-Krastanow growth mode, the QD states are located energetically below a quasicontinuum of delocalized states, which corresponds to the two-dimensional (2D) motion in a thin wetting layer (WL). Because of carrier-carrier and carrier-phonon scattering the WL is of great importance for QD devices if the excitation involves the quasicontinuum~\cite{Torben2004}. Computing the above many-particle processes requires the knowledge not only of the QD states but also of the spatially-extended WL states.
It is common that the latter are described by plane waves for their in-plane part~\cite{MUB02,CCH04}, thereby omitting the QDs' influence on the WL states. As a consequence, the QD and WL states are no longer orthogonal to each other, which potentially adulterates the QD-WL Coulomb coupling~\cite{Schneider01}. The orthogonalized plane wave (OPW) method is an attempt to remedy this shortcoming by a supplementary orthogonalization~\cite{BE92,FB99,Schneider01,Torben2004}. However, the resulting WL states are still approximated by plane waves, but with local modifications at the position of the QDs.  
Calculations without the above approximations exist for single QDs~\cite{MKT06,PBZ09,MLC18}, but the considered spatial domain around the QD is small, so the WL states are treated only locally and their properties are not in the focus.

In the field of quantum chaos~\cite{Gutz90,Stoeckmann00,Wimberger14}, plane wave-like energy eigenstates with a regular morphology appear in integrable systems. Examples of such analytically solvable systems are the rectangular and the circular billiard. The term billiard refers to a potential-free region enclosed by reflecting walls, in which a point mass moves along a straight line until it hits the boundary and bounces back with the angle of reflection equal to the angle of incidence. Most billiards are non-integrable and exhibit chaos, i.e. there exist trajectories with a sensitive dependence on initial conditions. Some geometries exhibit full chaos, i.e. almost all trajectories exhibit a sensitive dependence on initial conditions. The quantum properties of chaotic billiards have been studied experimentally in, e.g. optical microcavities (for a review see Ref.~\cite{CW15}) and ballistic semiconductor microstructures~\cite{MRWHG92}. The latter are sometimes also called QDs but are very different from the above self-assembled QDs, for instance because of the size (compared to the wavelength) and the lack of a WL.

\begin{figure}[tb]
\centering\includegraphics[width=0.67\linewidth]{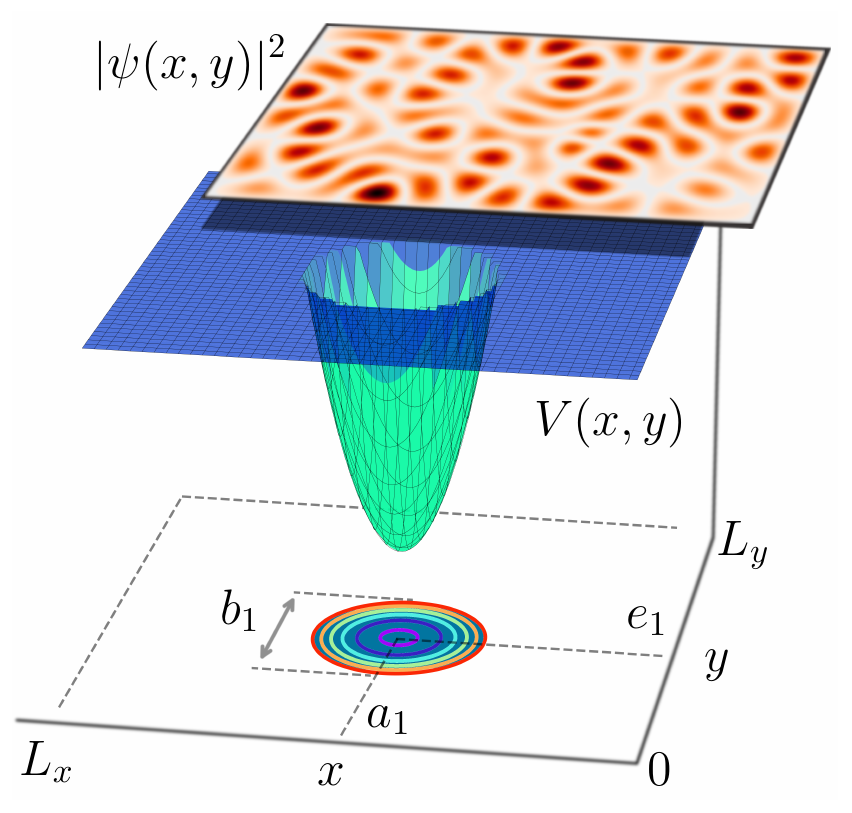}
	\caption{In-plane part of a wetting-layer state in a small rectangular Mesa with side lengths $L_x$ and $L_y$ containing a single quantum dot at $(x,y) = (a_1,e_1)$. The quantum-dot confinement potential $V$ is shown as a surface plot and as contours in the $x,y$-plane. Outside the quantum dot but inside the Mesa the potential~$V$ is zero.
}
	\label{fig:title}	
\end{figure}

The aim of this article is to demonstrate that even in the simple model of a rectangular-shaped WL with one or a few lens-shaped self-assembled QDs in the effective-mass approximation, the WL states are typically rather complex and do not resemble (orthogonalized) plane waves, as illustrated in Fig.~\ref{fig:title}. We use methods and concepts from the field of quantum chaos to study the complexity of the states.

The paper is organized as follows. Section~\ref{sec:model} introduces the model. Subsequently, Sec.~\ref{sec:numerics} describes the numerical method to calculate energy eigenstates and eigenvalues. The morphology of these states is qualitatively discussed in Sec.~\ref{sec:morph}. Section~\ref{sec:spectral_analysis} characterizes the system via spectral statistics. Sections~\ref{sec:amp_dist} and~\ref{sec:loc_dist} provide a quantitative study of the global and local spatial properties of the WL states. The results are compared to the OPW method in Sec.~\ref{sec:ortho}. Finally, the conclusion is given in Sec.~\ref{sec:conc}.

\section{The quantum-dot wetting-layer model}
\label{sec:model}
We describe shallow lens-shaped QDs sitting on a thin, quasi-2D WL in a one-band effective-mass approximation for the conduction-band electrons, ignoring the effects of band mixing, mass anisotropy, and spin-orbit coupling. A constant effective mass $M$ is used assuming that the QDs and the WL are of the same material and embedded in a different bulk material. For the thin WL we consider only the first subband exhibiting no nodes in $z$-direction. We determine the single-particle energy eigenstates in the envelope function approximation, and consider only the in-plane component $\psi(x,y)$ of the wave functions. The QD confinement is treated by a potential $V(x,y)$ which naturally also influences the spatially-extended states in the WL. The WL is considered as an area of finite size, also called Mesa (see, e.g.~\cite{GGS13}), which is chosen to be a region of rectangular shape $\Omega=[0,L_x]\times[0,L_y]$. Following Ref.~\cite{Schneider01} we describe the confinement in the WL by an infinitely high potential well in 2D.

It has been clearly demonstrated in theoretical~\cite{Wojs:96a,Riel08} and experimental~\cite{PSH98,TWP13} investigations that the single-particle energy spectrum of a shallow lens-shaped QD can be accurately reproduced by a truncated parabolic confinement potential with rotational symmetry in the $x,y$-plane. Thus the QD is a finite size version of an isotropic 2D harmonic oscillator. The energy levels of the isolated QD are therefore given by $\hbar \omega(n+1)$ with degree of degeneracy $n + 1$, where $n = 0,1,2,\ldots,n_{\text{max}}$. For a single dot the potential is depicted in Fig.~\ref{fig:title}.
Since the depositioning process in general leads to a spatial distribution of QDs on the WL we have to incorporate the $k$-th QD confinement potential to the overall WL-potential by summing over all $N_{\text{dot}}$ non-overlapping QDs
\begin{align}
&V(x,y)=\sum_{k=1}^{N_{\text{dot}}}V_k(x,y) \ , \label{eq:V}\\
&V_k(x,y)=
\text{min}\left\{d_k\left[(x-a_k)^{2}+(y-e_k)^{2}-\frac{b_k^2}{4}\right],0\right\},\label{eq:V2}
\end{align}
where the set of parameters $\{a_k,e_k,d_k,b_k\}$ characterizes the $k$-th QD confinement potential with its minimum at the coordinates $(a_k,e_k)$, width $b_k$, and depth $d_k b_k^2 /4$. 

Table~\ref{tab:para} summarizes the sets of parameters studied in the next sections. To avoid nongeneric behavior, we choose the aspect ratio $L_y/L_x$ to be an irrational number by using the reciprocal golden ratio $\Phi=(1+\sqrt 5)/2$. In our numerics we express the parameters $a_k$, $e_k$, $b_k$, $L_x$, and $L_y$ in nm and, using $\hbar =1 = M$, express energies in nm$^{-2}$.
First, we study the case of a single QD ($N_{\text{dot}}=1$) with three QD states. Three confined states are realistic for several applications such as QD lasing, see, e.g.~\cite{GWLJ06}. Within the model, for fixed $b_1$ the maximum number of QD states can be adjusted by the parameter $d_1$. Even though the situation with only one QD does not serve as a generic type of a semiconductor with QD densities around $10^{10}\text{cm}^{-2}$ it is experimentally feasible in a Mesa, as in Ref.~\cite{GGS13}, and offers the opportunity to study the impact on the system properties on the most simplified level. The first sample system is referred to as~$\mathcal A(1)$. The QD position $(a_1,e_1)$ is chosen such that the QD is not exactly at the center to avoid nongeneric effects due to symmetry. 
Leaving $d_1$ and~$b_1$ unchanged we extend the area~$\Omega$ by multiplying $L_x$ and $L_y$ of system $\mathcal A(1)$ by $\test = 2$, $3$, and $4$. Also the QD position is changed by the same factor such that $L_x/a_1$ and $L_y/e_1$ remain the same for all $t$. In the same order, these systems are labeled by $\mathcal A(2)$, $\mathcal A(3)$, and $\mathcal A(4)$. Moreover, another system containing an ensemble of four identical, non-overlapping QDs is denoted by~$\mathcal B$. System $\mathcal C$ possesses the same parameter settings as system $\mathcal A(1)$ but with an adjusted parameter $d_1$ reducing the number of QD states to one. 
\setlength{\tabcolsep}{4.2pt} 
\begin{table}[ht]
\centering\begin{tabular}{c c c c c c c c c c}
	\hline\hline
	System & $N_{\text{dot}}$ &$k$ & $a_k$ & $e_k$ & $d_k$ & $b_k$ & $L_x$ & $L_y$\\
	\hline
	$\mathcal A(\test)$ & $1$ & $1$ & $4.3\test$ & $6.3\test$ & $2.9$ & $2.5$ & $10\test$ & $10 \test\Phi$ \\
	$\mathcal B$ & $4$ & $1$ & $11.575$ & $18.698$ & $2.9$ & $2.5$ & $34$ & $15\Phi$ \\
	& & $2$ & $24.135$ & $13.708$ &  $2.9$ & $2.5$  &  &  \\
	& & $3$ & $25.827$ & $9.446$ &  $2.9$ & $2.5$  &  &  \\
	& & $4$ & $5.562$ & $3.297$ &  $2.9$ & $2.5$  &  &  \\
	$\mathcal C$ & $1$ & $1$ & $4.3$ & $6.3$ & $1.1$ & $2.5$ & $10$ & $10\Phi$  \\
	\hline\hline
\end{tabular}
	\caption{Parameter sets of different QD-WL systems. $a_k$, $e_k$, $b_k$, $L_x$, and $L_y$ are given in nm. $d_k$ is given in nm$^{-4}$.}
	\label{tab:para}
\end{table}

The restriction to the first subband implies a cut-off energy $\Eco$ below which the system behaves quasi-2D. This is analog to the case of microwave billiards, which are popular experimental systems in the field of quantum chaos, see, e.g.~\cite{Stoeckmann00}. The cut-off energy is determined by the condition that the second subband appears which happens in our case at $\Eco = 3\pi^2\hbar^2/(2M\dWL^2)$ where $\dWL$ is the thickness of the WL. Assuming a thin WL with $\dWL = 0.4\,$nm we get $\Eco \approx 92\,$nm$^{-2}$. In the following we will consider only energy levels below this cut-off energy.

We mention that the model introduced above is related to the model in Ref.~\cite{MKT06} with, in the context of the present paper, two important differences: (i) in Ref.~\cite{MKT06} only a single QD is studied and (ii) this dot is sitting on the center of a circular-shaped WL. This composed system is nongeneric as it possesses a rotational symmetry. The resulting integrability is convenient for computations but it leaves no room for a complex morphology of energy eigenstates. In strong contrast, our model is nonintegrable and allows to study generic features of QD-WL systems. 

At first glance at the lower part of Fig.~\ref{fig:title} there is also a superficial similarity to the Sinai billiard~\cite{Sinai70}, a square billiard with a circular wall located at its center, and the {\v Seba} billiard~\cite{Seba90}, a rectangular billiard containing a point scatterer (see also the application to ballistic QDs in Ref.~\cite{RH10b}). However, our QD-WL model is very different as the potential is attracting, nonuniform, continuous, and finite. 

Finally, our model is related to the quasi-2D microwave billiard with randomly placed spherical caps which has been studied in Ref.~\cite{BMF13} in the context of branch flows in random potentials. These caps remind strongly on our lens-shaped QDs positioned {\it on} the WL but the caps are placed {\it inside} the resonator. The latter results in repulsive local potentials whereas in our case the local potentials are attractive. The energy eigenstates and eigenvalues have not been discussed in Ref.~\cite{BMF13}.

\section{Computation of eigenstates and eigenvalues}
\label{sec:numerics}
We consider the quantum-mechanical eigenvalue problem
\begin{align}
{\hat H}\psi_i(x,y) = E_i\psi_i(x,y)
\label{eq3}
\end{align}
with the single-particle Hamiltonian 
\begin{align}
{\hat H}=-\frac{\hbar^2}{2M}\Delta_{x,y}+V(x,y)
\label{eq:Ham}
\end{align}
and the eigenstates $\psi_i$ fulfilling zero Dirichlet boundary conditions on the WL boundary $\partial \Omega$. We solve it numerically by expressing ${\hat H}$ in a suitable orthonormal basis $\{\varphi_n\}_{}^{}$ and diagonalizing the associated Hamiltonian matrix $H=(H_{n,m})$. This procedure supplies the energy eigenvalues~$E_i$ which, due to considering a finite region $\Omega$, are discretized, and the coefficients $c_{n}^{(i)}=\langle\varphi_n\vert\psi_i\rangle_{}$ required for the series expansion of the eigenstates 
\begin{align}
\psi_i(x,y) = \sum_{n}^{} c_{n}^{(i)}\,\varphi_{n}(x,y) \ .
\label{eq:expansion}
\end{align}
Here, it is convenient to choose the eigenbasis $\{\varphi_n\}$ of $\hat H_0$ resulting from removing the QD potential from the Hamiltonian in Eq.~(\ref{eq:Ham}). According to the boundary conditions the basis is given by the plane wave-like (checkerboard-like) states
\begin{align}
\varphi_{n}(x,y) = \langle x,y\vert\varphi_n\rangle &= \frac{2}{\sqrt{L_xL_y}}\sin(k_{x}x)\sin(k_{y}y) 
\label{eq:planewaves}
\end{align}
with wave numbers $k_{x} = n_x\pi/L_x$, $k_{y} = n_y\pi/L_y$, and $n_x, n_y \in \mathbb{N}^{+}$. The multi-index $n=\{n_x,n_y\}$ can be assembled by using an appropriate pairing function facilitating a bijectively mapping of two integers onto one and thus providing a certain ordering scheme. Because $n_x, n_y \in \mathbb{N}^{+}$ we are using a bijective map $h: \mathbb N^{+} \times \mathbb N^{+}\rightarrow \mathbb N^{+}$ introduced by Hopcroft and Ullman~\cite{Hopull79}. 

By construction, the matrix elements of the first term in Eq.~(\ref{eq:Ham}), recognized as $\hat H_0$, can be evaluated analytically
\begin{align}
\langle\varphi_n \vert \hat H_0 \vert \varphi_m\rangle
=E_{n_x,n_y} \delta_{n_x,m_x}\delta_{n_y,m_y}
\label{eq12}
\end{align}
with $E_{n_x,n_y} = \hbar^2\left(k_{x}^2 + k_{y}^2\right)/2M$. The matrix elements of the second term in Eq.~(\ref{eq:Ham}) can be written as
\begin{align}
\langle\varphi_n \vert V \vert \varphi_m\rangle
=& \sum_{k=1}^{N_{\text{dot}}} d_k  \iint \limits_{B_k}^{} dy\, dx\ \varphi_{n}^*(x,y)\varphi_{m}(x,y)\,\nonumber \\
&\ \ \ \ \ \ \times 
\left[(x-a_k)^2+(y-e_k)^2-\frac{b_k^2}{4}\right],
\label{eq:doubleintegral}
\end{align}
with $B_k$ constituting the region where $V_k\neq 0$; cf. Fig.~\ref{fig:title}. The numerical evaluation of the double integral~(\ref{eq:doubleintegral}) is done in local polar coordinates by using \textit{scipy.integrate.nquad} from the SciPy package~\cite{Jones01}. The upper bound on the number of subintervals used in the adaptive algorithm has been set to 100 to ensure convergence.
For the computation of eigenvectors and eigenvalues of the real symmetric matrices we use \textit{numpy.linalg.eig} and \textit{numpy.linalg.eigvalsh} from the same package with default settings. The resulting eigenstates and eigenvalues are ordered such that $E_1\leq E_2\leq \ldots$.

\section{Morphology of energy eigenstates}
\label{sec:morph}
\begin{figure}[tb]
\centering\includegraphics[width=1.0\linewidth]{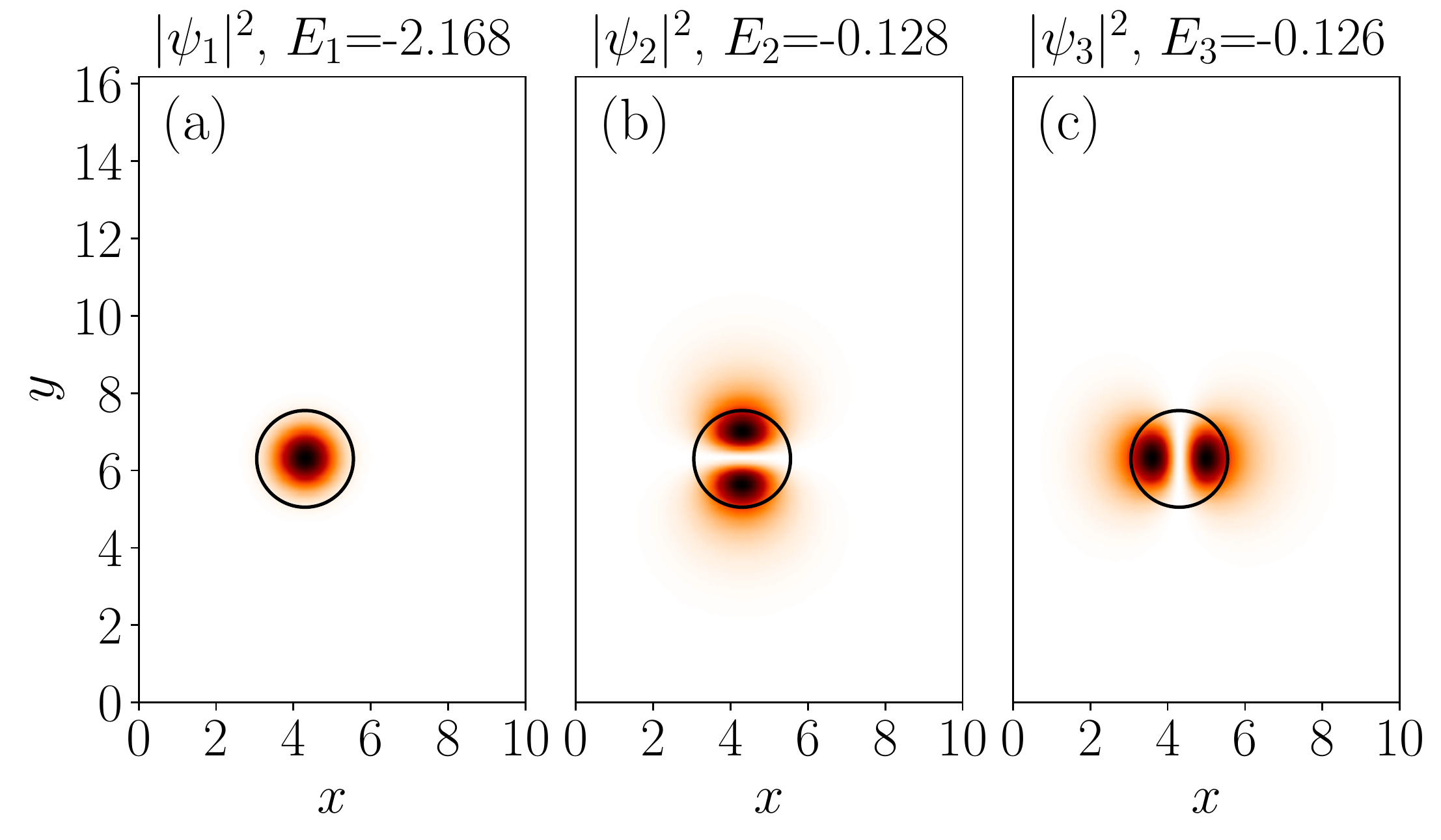}
	\caption{Probability density $\vert\psi_i(x,y)\vert^2$ of the three QD states [(a) $s$-shell, (b)-(c) $p$-shells] of single-QD system $\mathcal A(1)$ numerically obtained for a series expansion in Eq.~(\ref{eq:expansion}) containing $3000$ terms. Here the boundary of the QD is indicated as a circle. The colormap ranges from zero (white) over red up to the highest value of $\vert\psi\vert^2$ (black). $x$ and $y$ are measured in nm and energies $E_i$ are measured in nm$^{-2}$.}
	\label{fig:ground_states_1}	
\end{figure}

This section provides a qualitative discussion of the numerically obtained eigenstates. For a quantitative analysis we refer to Secs.~\ref{sec:amp_dist} and~\ref{sec:loc_dist}. As representatives we choose systems $\mathcal A(1)$ and $\mathcal B$, cf. Table~\ref{tab:para}. 
For the single-QD system $\mathcal A(1)$ the numerically computed QD states are shown in Fig.~\ref{fig:ground_states_1}. We define the QD states as those states having negative energy; see Eqs.~(\ref{eq:V})-(\ref{eq:V2}). The QD states are localized in the QD and obey the expected behavior of an isotropic 2D harmonic oscillator. The ground state ($s$-shell) is concentrated in the center of the QD and its energy is well approximated by a 2D harmonic oscillator, shifted by $V_{\text{min}}=-d_1b_1^2/4$. The next two consecutive states ($p$-shells) possess the typical appearance with probability densities leaking into the classically prohibited region. The eigenenergies are slightly splitted due to the presence of the WL boundary.

\begin{figure*}[tb]
\centering\includegraphics[width=0.73\linewidth]{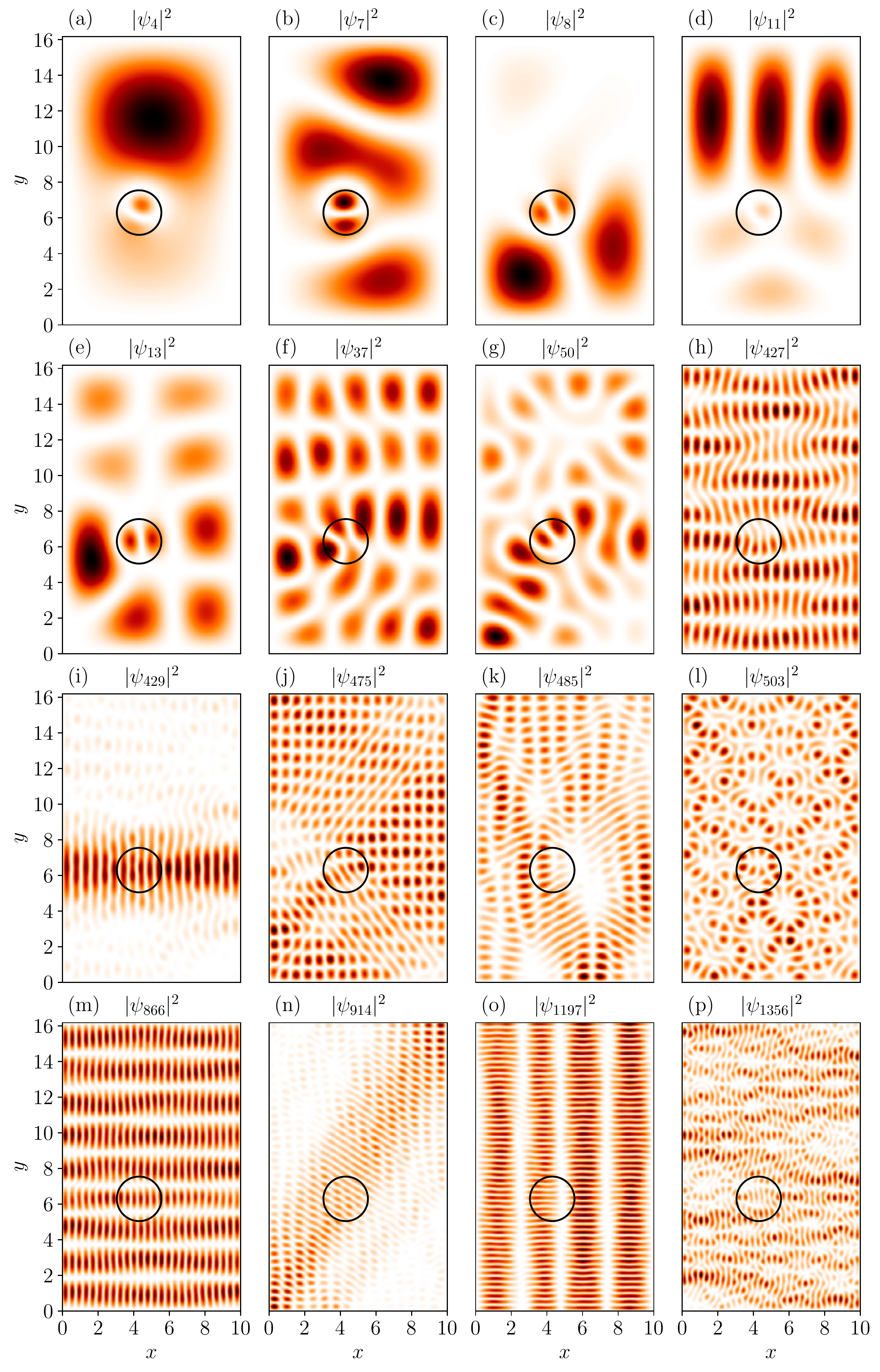} 
	\caption{Probability density of some WL states in the single-QD system $\mathcal A(1)$. The series expansions contain $3000$ terms each. Assignments of colors are the same as in Fig.~\ref{fig:ground_states_1} and the QD is indicated as a circle. $x$ and $y$ are measured in nm.}
	\label{fig:unbound_1}	
\end{figure*}

WL states have nonnegative energies and are therefore spatially extended. Figure~\ref{fig:unbound_1} shows that these states are strongly affected by the presence of the QD. While some states like $\psi_{8},\psi_{50}$, and $\psi_{1356}$ display an irregular or nonuniformly spatial structure, some others like $\psi_{475}, \psi_{866}$, and $\psi_{1197}$ display regular and almost uniform checkerboard-like densities. Occasionally, for system $\mathcal A(1)$ we also find characteristic states like $\psi_{429}$ which might be classified as \quoting{bouncing-ball states}~\cite{BSS97}.

Furthermore, by taking a closer look at the WL states in Fig.~\ref{fig:unbound_1} we notice that some of them exhibit a higher probability in the region of the QD, e.g. $\psi_{7}, \psi_{8}$, and $\psi_{13}$. Some others like $\psi_{11}$ and $\psi_{37}$ show a diminished probability. The latter states seem to tend to \quoting{elude} the QD. The examination of this effect on a statistical level is the subject of Sec.~\ref{sec:loc_dist}.
  
\begin{figure*}[tb]
\centering\includegraphics[width=0.85\linewidth]{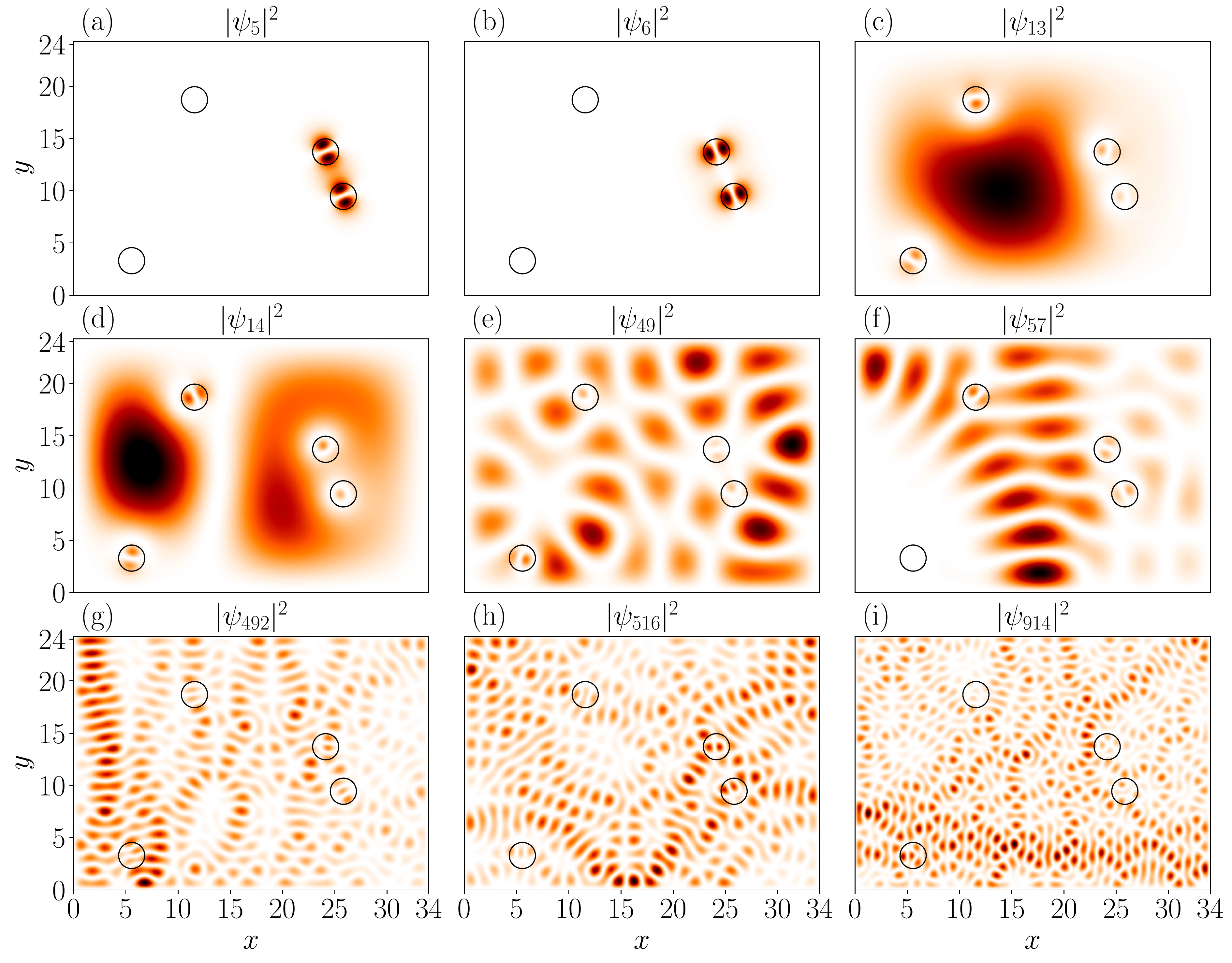}
	\caption{Probability density of some energy eigenstates in the 4-QD system $\mathcal B$. Here the series expansions contain $4000$ terms. Assignments of colors are the same as in Fig.~\ref{fig:ground_states_1} and QDs are indicated as circles. $x$ and $y$ are measured in nm.}
	\label{fig:states_2}	
\end{figure*}

Whereas the WL states of system $\mathcal A(1)$ constitute a set of several types, for system~$\mathcal B$ Fig.~\ref{fig:states_2} shows that increasing the number of QDs implies a loss of diversity with respect to the spatial morphology. Checkerboard-like states have disappeared and generally the spatial structure of the higher-excited states looks more irregular.
Additionally, due to the QD configuration in system $\mathcal B$ we come across the opportunity to observe coupled QD states when the inter-dot distance is small enough, like in Figs.~\ref{fig:states_2}(a) and (b).

To conclude the qualitative discussion, we can say that the morphology of the WL states in this simple QD-WL model is much more complex than that of plane waves.

\section{Spectral analysis}
\label{sec:spectral_analysis}
\begin{figure}[tb]
\centering\includegraphics[width=1.0\linewidth]{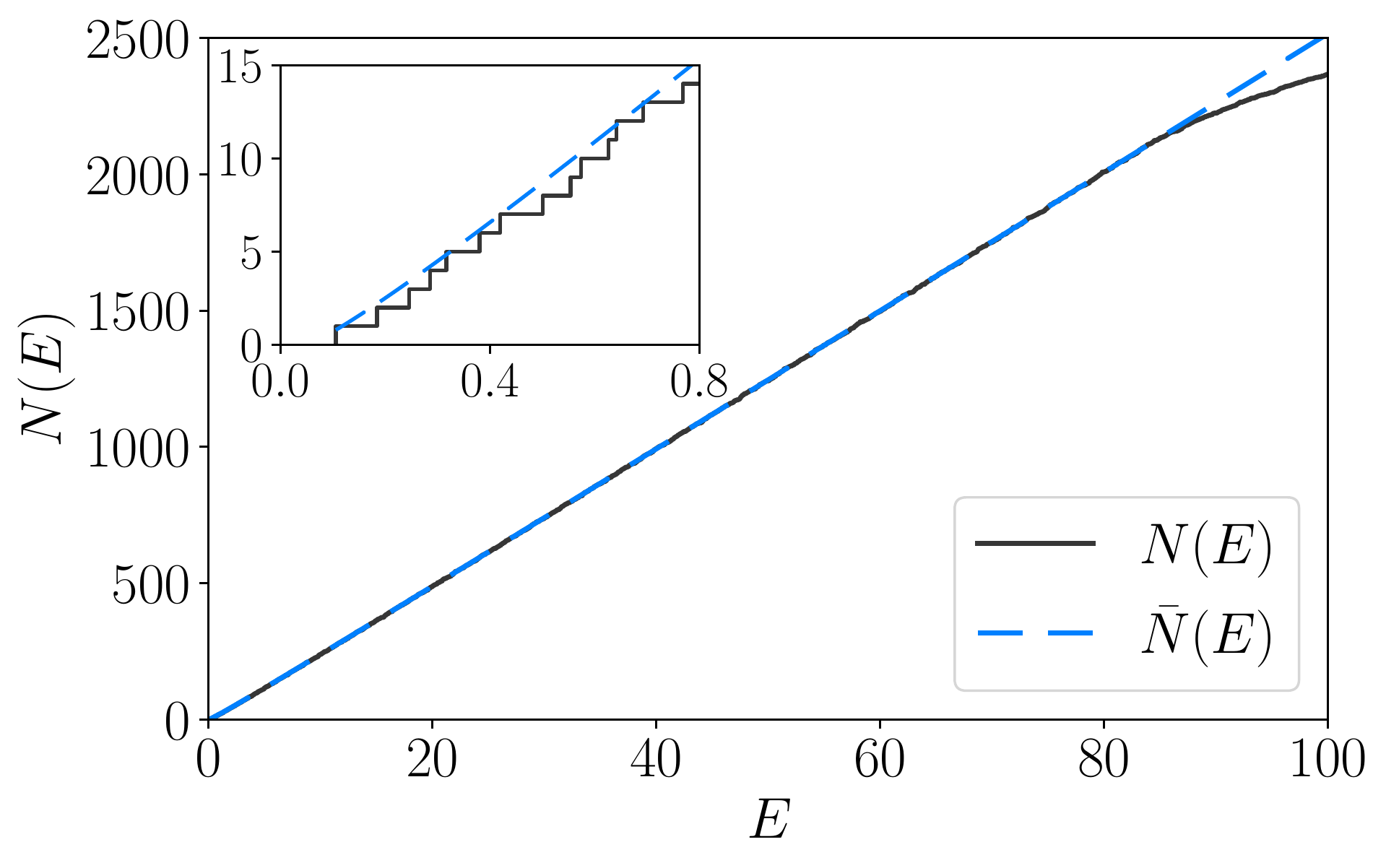}
	\caption{Integrated level density $N(E)$ (number of levels up to the energy $E$) and its smooth part ${\bar N}(E)$ according to Weyl's law in Eq.~(\ref{eq:Weyl}) for system $\mathcal A(1)$. The inset magnifies the low-energy region. The energy $E$ is given in nm$^{-2}$.} 
	\label{fig:unfolding}	
\end{figure}

Here, we investigate the information contained in the energy spectrum utilizing statistical measures developed in the field of quantum chaos. 
Two of them are the so-called level spacing or nearest-neighbor spacing distribution $P(s)$ which reveals the short-range correlations of the energy levels and secondly the number variance $\Sigma^2(L)$ revealing the long-range level correlations, see, e.g.~\cite{Stoeckmann00,Wimberger14}. Both quantities are dealing with the fluctuations within a discrete level sequence $\{E_i\}$ with $E_1\leq E_2\leq \ldots$. We start with the level density $\rho(E) = \sum_{i}^{} \delta(E -E_i)$ and the integrated level density 
\begin{align}
N(E) = \int\limits_{-\infty}^{E}dE'\rho(E')=\sum_{i}^{}\Theta(E-E_i)
\end{align}
giving the number of levels up to the energy $E$. Here $\Theta$ denotes the Heaviside step function.
$N(E)$ can be segregated into two parts, a smooth part ${\bar N}(E)$ and a fluctuating one
\begin{align}
N(E) = {\bar N}(E)+N_{\text{fluc}}(E) \ .
\end{align}
Determining the smooth part ${\bar N}(E)$ is in general a nontrivial task~\cite{GMRR}.
However, for 2D quantum billiards of arbitrary shape with area $A$ and circumference $L$, the smooth part of the integrated level density, for not too small energies $E$, can be very well approximated by the generalized Weyl's law~\cite{Gutz90}
\begin{align}
{\bar N}(E) \approx \frac{A}{4\pi}\frac{2ME}{\hbar^2} - \frac{ L}{4\pi}\sqrt{\frac{2ME}{\hbar^2}}\ .
\label{eq:Weyl}
\end{align}
We assume (and verify below) that this law can be also applied to our QD-WL model with $A=L_x L_y$ and ${L} = 2(L_x+L_y)$ if the levels of the QD states are omitted. This is justified as Weyl's law is only valid for not too small energies and, moreover, a few events are insignificant to the overall statistics. 
The validity of Weyl's law is demonstrated for system $\mathcal A(1)$ in Fig.~\ref{fig:unfolding}. The spectrum is calculated for a Hamiltonian matrix of size $3000\times 3000$. Up to the $2100$th energy level Weyl's law constitutes a marvelous approximation to the integrated level density. Deviations beyond this level can be attributed to truncating the Hilbert space.

\begin{figure}[tb]
\centering\includegraphics[width=1.0\linewidth]{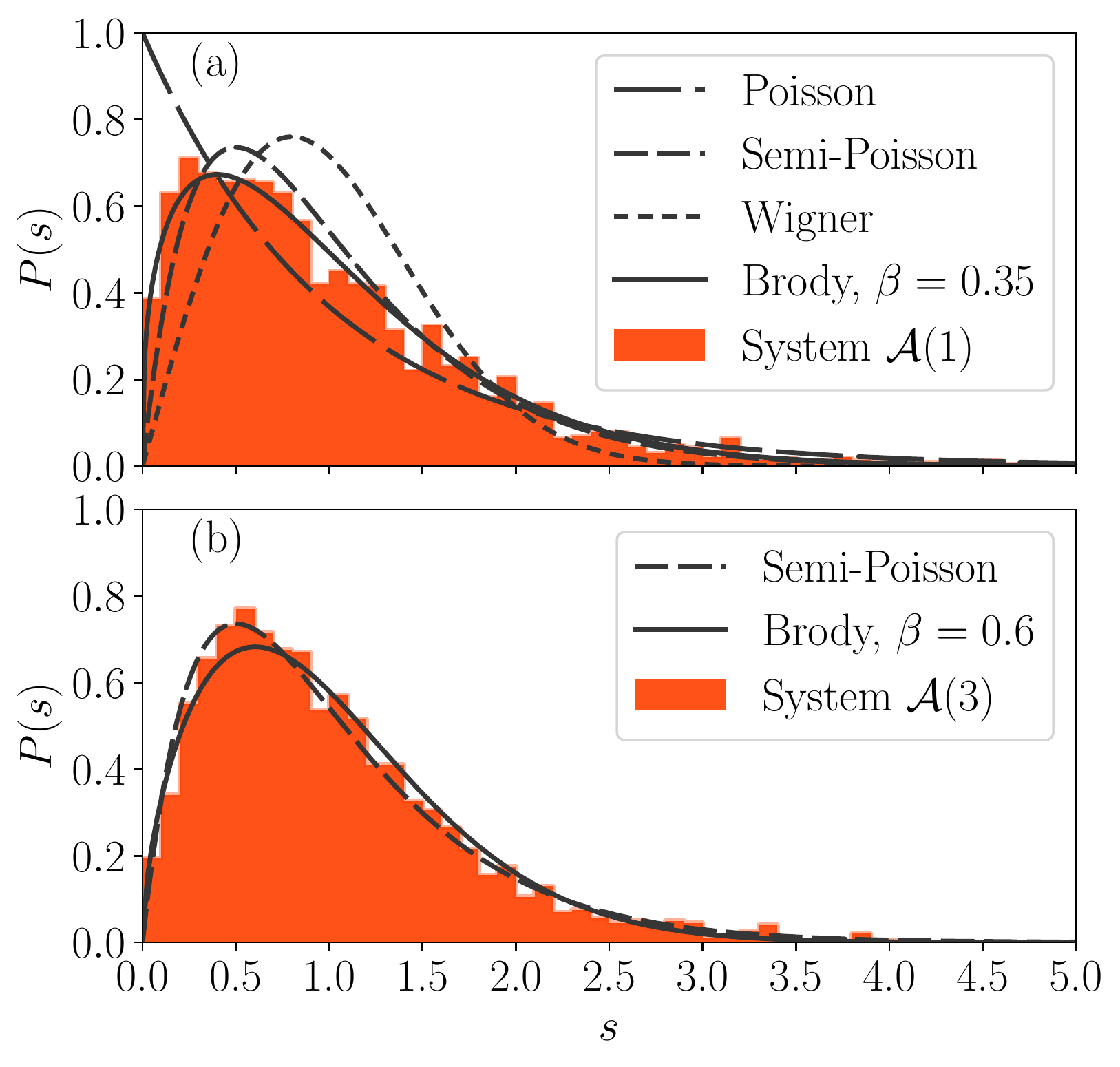}
	\caption{Probability distribution $P(s)$ of the (dimensionless) nearest-neighbor spacing $s$ of the single-QD systems $\mathcal A(1)$ in (a) and $\mathcal A(3)$ in (b) for the first $2000$ WL states. For comparison the Poisson statistics [Eq.~(\ref{eq:Poisson})], Semi-Poisson statistics [Eq.~(\ref{eq:SP})], Wigner surmise [Eq.~(\ref{eq:Wigner})], and the Brody distribution [Eq.~(\ref{eq:Brody})] is shown.}
	\label{fig:level_3}	
\end{figure}

With the aid of ${\bar N}(E)$, the spectrum $\{E_i\}$ can be mapped onto the so-called unfolded spectrum $\{e_i\}$ through $e_i = {\bar N}(E_i)$. This procedure separates the local level fluctuations from an overall energy dependence. 
Owing to the unfolding the sequence of nearest-neighbor energy spacings $\{s_i = e_{i+1}-e_i\}$  constitutes a set for which the mean level spacing $\langle s\rangle\approx 1$. The corresponding probability distribution is called $P(s)$. 
Uncorrelated energy levels follow the Poisson statistics with nearest-neighbor spacing distribution
\begin{align}
P_{\text{P}}(s)= e^{-s} \ .
\label{eq:Poisson}
\end{align}
In the field of quantum chaos the Poisson statistics describe the level statistics of generic integrable systems. This is also true for the special case of a rectangular billiard with a generic aspect ratio~\cite{Marklof98} which is nothing else than our QD-WL model without QDs having plane wave-like energy eigenstates~(\ref{eq:planewaves}). 
In the presence of a QD deviations from Poisson statistics can be observed, e.g. in Fig.~\ref{fig:level_3}(a) for system $\mathcal A(1)$. These deviations therefore reveal correlations between adjacent energy levels, i.e. short-range level correlations. In particular, $P(s)$ is significantly reduced for small values of $s$. Hence, it is less probable to observe level crossings when a parameter is varied. 
\begin{figure*}[tb]
\centering\includegraphics[width=1.0\linewidth]{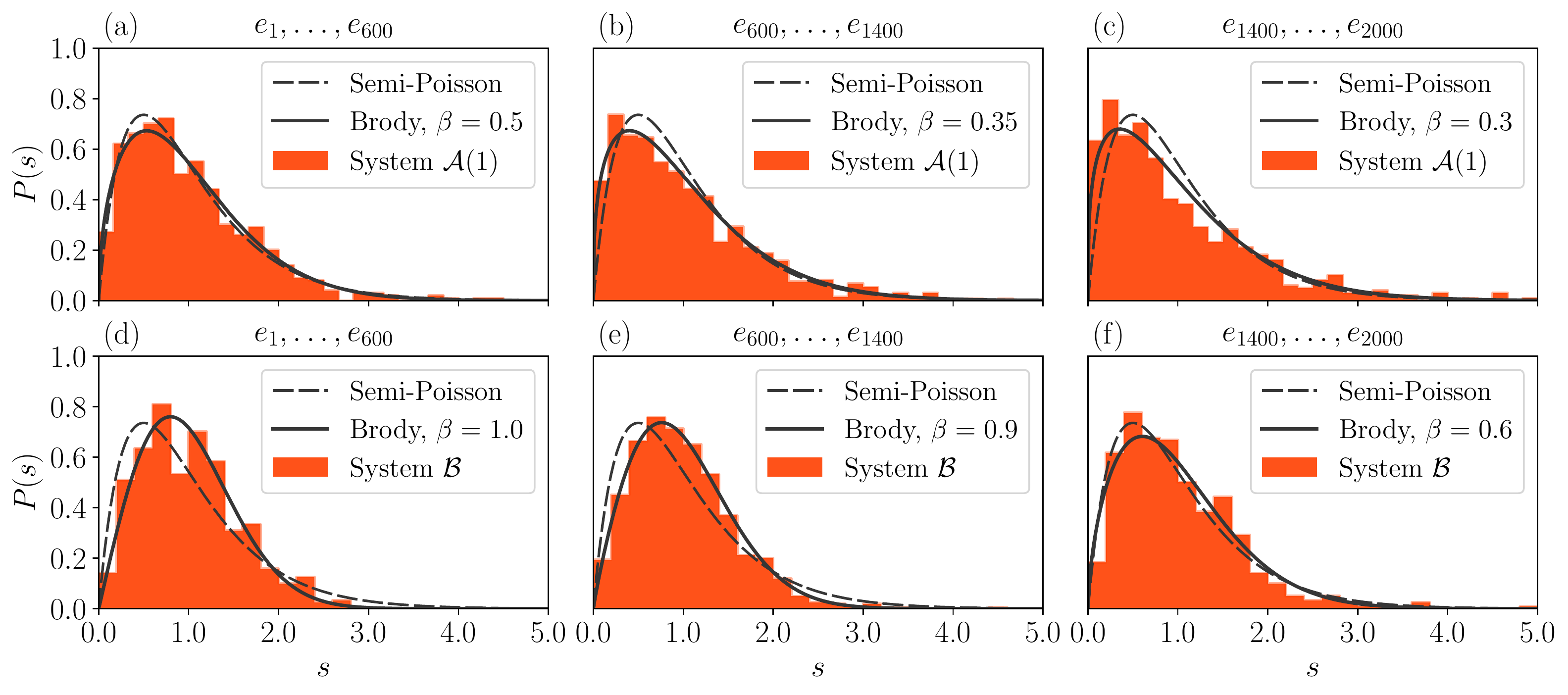}
	\caption{Probability distribution $P(s)$ of the (dimensionless) nearest-neighbor spacing $s$ of single-QD system $\mathcal A(1)$ [(a)-(c)] and $4$-QD system $\mathcal B$ [(d)-(f)] in different energy windows.}
	\label{fig:level_6}	
\end{figure*}
%

Figure~\ref{fig:level_3}(a) also shows that $P(s)$ cannot be described by the Gaussian orthogonal ensemble (GOE) of random-matrix theory which applies to classically fully chaotic systems with time-reversal symmetry. The nearest-neighbor spacing distribution of the GOE is well approximated by the Wigner surmise (see, e.g.~\cite{Wimberger14})
\begin{align}
P_{\text{W}}(s)=\frac{\pi}{2}s\, e^{-\pi s^2/4} \ .
\label{eq:Wigner}
\end{align}

A better agreement for system $\mathcal A(1)$ is achieved with the so-called Semi-Poisson statistics~\cite{Wiersig02,BGS01b} with nearest-neighbor spacing distribution
\begin{align}
P_{\text{SP}}(s) = 4s\, e^{-2s} \ .
\label{eq:SP}
\end{align}
Like the Wigner surmise $P_{\text{SP}}(s)$ shows a linear behavior at small $s$. But for large $s$ it exhibits an exponential fall-off like in the Poisson case. The Semi-Poisson statistics is a  model for intermediate spectral statistics~\cite{BGS01b} and appears, e.g. in pseudointegrable systems~\cite{Wiersig02,GW03,WC03} and multi-walled carbon nanotubes~\cite{AKWC02}. Furthermore, it has also been controversially discussed in the context of the {\v Seba} billiard~\cite{TKS10}. 

Next, we check the Brody distribution~\cite{BFF81}
\begin{align}
P_{\text{B}}(s,\beta) = (\beta + 1)b s^\beta e^{-bs^{\beta+1}}
\label{eq:Brody}
\end{align}
with 
\begin{align}
b = \left[\Gamma\left(\frac{\beta+2}{\beta+1}\right)\right]^{\beta+1}
\end{align}
and the gamma function $\Gamma$. The Brody distribution is a heuristic interpolation between Wigner surmise ($\beta = 1$) and Poisson statistics ($\beta = 0$), see Ref.~\cite{Stoeckmann00}. Figure~\ref{fig:level_3}(a) shows a very good agreement with the numerical data for $\beta = 0.35$, which is even slightly better than for the Semi-Poisson statistics. 

We have also tested the Berry-Robnik distribution~\cite{BR84}
\begin{align}
P_{\text{BR}}(s,\beta) & = \alpha^2e^{-\alpha s}\text{erfc}\left(\frac{\sqrt{\pi}}{2}\beta s\right)&\nonumber\\
 &\ \ \ \ +\left(2\alpha\beta+\frac{\pi}{2}\beta^3s\right) e^{-\alpha s-\pi \beta^2s^2/4}\ ,
\label{eq:BR}
\end{align}
with the complementary error function $\text{erfc}$ and $\alpha=1-\beta$. This distribution also interpolates between Wigner surmise ($\beta = 1$) and Poisson statistics ($\beta = 0$). 
In contrast to the Brody distribution, it has a theoretical foundation for systems with a mixed classical phase space where regular/integrable and chaotic dynamics coexist. 
The spectra associated with the corresponding regions in phase space are assumed to be statistically independent, and their mean level spacing is determined by the size of the regions.
Even though our QD-WL model should be in the generic class of systems with mixed phase space, the Berry-Robnik distribution does not give a satisfactory fit to the data for $s\leq 1$ (not shown). This failure of the Berry-Robnik distribution is consistent with other numerical studies~\cite{PR93,ZCZ10}. It results from neglecting dynamical tunneling between regular and chaotic regions~\cite{BKL11}.

We observe that the agreement of the numerical data and the Brody distribution remains reasonable when the WL area is increased by going through the systems $\mathcal A(\test)$ with $\test = 2,3,4$ (keeping the size of the Hamiltonian matrix constant), see Table~\ref{tab:para}. For $\test=3$ (not shown for $\test = 2$ and $4$) this can be clearly observed in Fig.~\ref{fig:level_3}(b). It can also be seen that in this case the Semi-Poisson statistics gives a slightly better agreement. 

Following the discussion in Refs.~\cite{AKWC02,TKS10} we also consider the spectral statistics on various energy windows. Dividing the overall spectrum of $2000$ energies into smaller subsets reveals that $P(s)$ of system $\mathcal A(1)$ gets different contributions from separate energy windows as shown in Figs.~\ref{fig:level_6}(a)-(c). In the lowest energy window $P(s)$ is equally well described by the Semi-Poisson statistics and by a Brody distribution with fitting parameter $\beta = 0.5$. For increasing energies the distribution is slightly better described by Brody distributions with $\beta$ decreasing to $0.35$ and $0.3$, i.e. the distribution slowly shifts closer to the Poisson statistics. This is reasonable as the influence of the QD potential is reduced for high energies. If the QD potential can be ignored then the system is effectively a rectangular billiard obeying Poisson statistics. 

The tendency to Poisson statistics for higher energies is also observable when more than one QD is present. Figures~\ref{fig:level_6}(d)-(f) show the case of four QDs in system~$\mathcal B$ with altogether twelve QD states using a $4000\times 4000$ Hamiltonian matrix. For low energies, the nearest-neighbor spacing distributions [Fig.~\ref{fig:level_6}(d)] can be characterized by the Brody distribution with $\beta = 1$ which equals the Wigner surmise. By increasing the energy a slow transition to $\beta = 0.9$ [Fig.~\ref{fig:level_6}(e)] and $\beta = 0.6$ [Fig.~\ref{fig:level_6}(f)] is observed. In the high-energy regime also the Semi-Poisson statistics describes the numerical data very well. 
Note that system~$\mathcal B$ is significantly closer to the statistics of random-matrix theory than system~$\mathcal A(1)$ even though the QD density differs just by a factor of $1.28$. This is consistent with our findings on the energy eigenstates in Sec.~\ref{sec:morph}, see Figs.~\ref{fig:unbound_1} and~\ref{fig:states_2}.

To study long-range level correlations we consider the number variance 
\begin{align}
\Sigma^2(L) =\langle (n(L,e)-L)^2\rangle \ ,
\label{eq:nv}
\end{align}
measuring the local variance of the number $n(L,e)=N(e+L/2)-N(e-L/2)$ of energy levels in the interval $\left[e-L/2,e+L/2\right]$. The bracket $\langle\ldots\rangle$ denotes the average over all spectral positions $e$. 
For the Poisson statistics one gets $\Sigma^2_{\text{P}}(L) =L$, for the Semi-Poisson statistics~\cite{Wiersig02,BGS99}
\begin{align}
\Sigma^2_{\text{SP}}(L) = \frac{L}{2}+\frac{1}{8}(1-e^{-4L}) \ ,
\label{eq:SPnv}
\end{align}
and for the GOE (see, e.g.~\cite{Wimberger14})
\begin{align}
\Sigma^2_{\text{GOE}}(L) = \frac{2}{\pi^2}\left[\ln(2\pi L)+\gamma + 1- \frac{\pi^2}{8}\right]+\mathcal O(L^{-1}) \ ,
\label{eq:GOEnv}
\end{align}
with the Euler constant $\gamma\approx 0.57722$. 

\begin{figure}[tb]
\centering\includegraphics[width=1.0\linewidth]{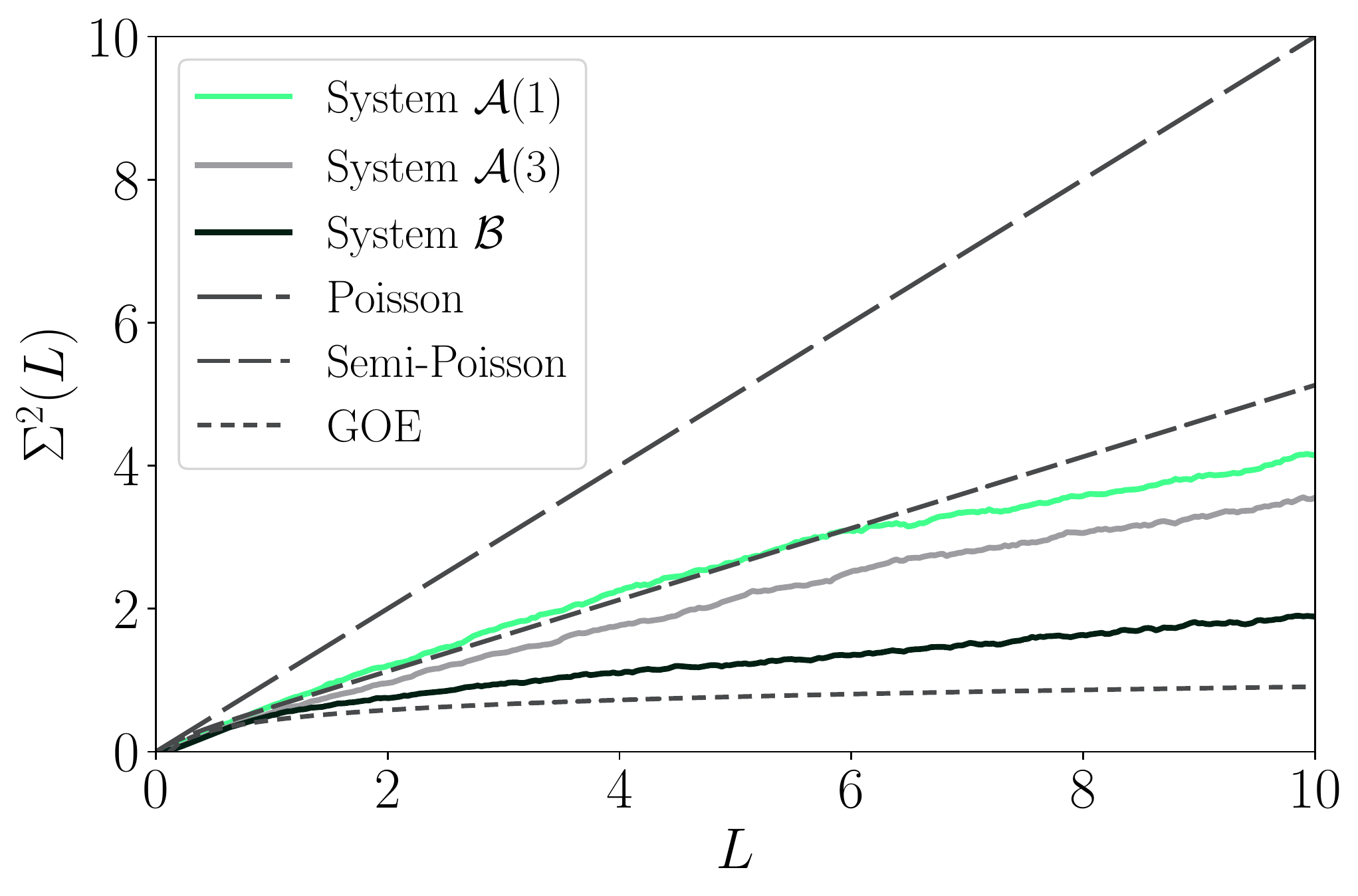}
	\caption{Number variance~(\ref{eq:nv}) of systems $\mathcal A(1)$, $\mathcal A(3)$, and $\mathcal B$ considering $2000$ levels each. For comparison the Poisson statistics, Semi-Poisson statistics [Eq.~(\ref{eq:SPnv})], and GOE [Eq.~(\ref{eq:GOEnv})] is shown.}
	\label{fig:number_variance}	
\end{figure}

In the case of the single-QD system $\mathcal A(1)$ Fig.~\ref{fig:number_variance} shows a remarkable agreement with the number variance of the Semi-Poisson statistics in Eq.~(\ref{eq:SPnv}). For $L > 6$ we observe deviations which might be the generic saturation behavior due to a finite number of energy levels used, see, e.g. Ref.~\cite{BSS95}.
For system $\mathcal A(3)$ with one QD on a three times larger WL, Fig.~\ref{fig:number_variance} shows a shift towards the GOE in Eq.~(\ref{eq:GOEnv}). For the 4-QD system $\mathcal B$ we find that $\Sigma^2(L)$ is even closer to $\Sigma^2_{\text{GOE}}(L)$.
To summarize, the short- and long-range level correlations are intermediate between Poisson statistics and GOE with a tendency towards GOE when the size of the WL or the number of QDs is increased. This shows that the WL states in our model show features known from irregular energy eigenstates in nonintegrable systems. 

\section{Amplitude distribution}
\label{sec:amp_dist}
After the qualitative discussion on the morphology of the WL states in Sec.~\ref{sec:morph}, we now provide quantitative measures for the observed complexity. 
According to Berry~\cite{Berry77}, the (real-valued) amplitude~$\psi$ of an individual \quoting{chaotic} or irregular wave function behaves like a Gaussian random variable. Hence, the probability of finding a value of it at any arbitrary given point in a region of area $A$ is~\cite{MK88}
\begin{align}
P(\psi)=\left(\frac{A}{2\pi}\right)^{1/2}e^{-A\psi^2/2} \ .
\label{eq:ad}
\end{align}
To determine $P(\psi)$ for a given WL state we evaluate its wave function $\psi(x,y)$ at $128\times128$ points in the $x,y$-plane. The outermost points of the WL region~$\Omega$ are excluded to avoid an artificial peak at $\psi=0$ due to the zero Dirichlet boundary conditions. Then $P(\psi)$ is generated as a histogram with $90$ bins in $[-0.5,\ldots,0.5]$, normalized to unity. 

\begin{figure}[tb]
\centering\includegraphics[width=1.0\linewidth]{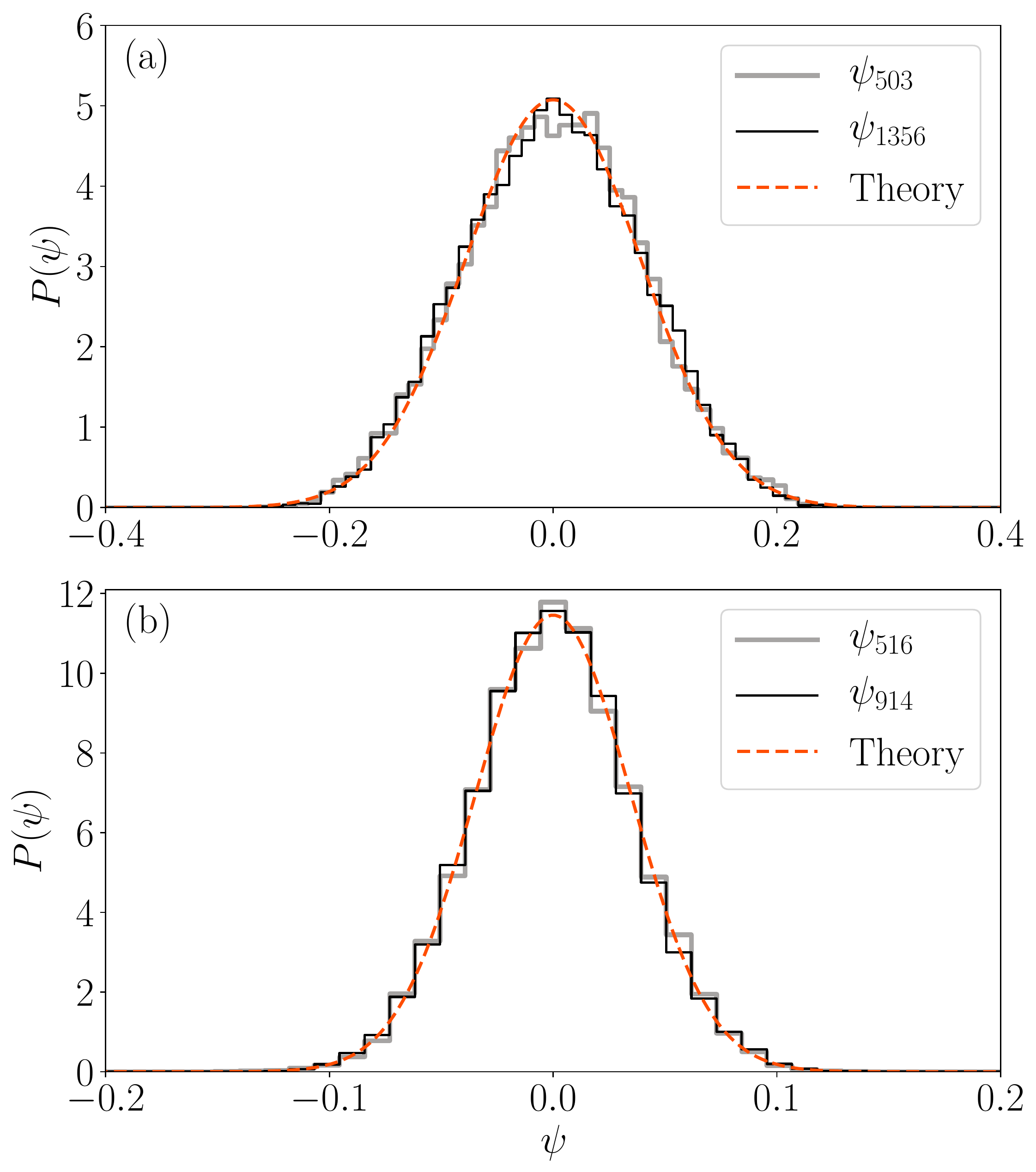}
	\caption{Probability amplitude distribution $P(\psi)$ for two WL states of (a) single-QD system $\mathcal A(1)$ and (b) 4-QD system $\mathcal B$ respectively (histograms); cf. Figs.~\ref{fig:unbound_1} and \ref{fig:states_2} for the morphology of the states. The dashed curve is the theoretical result for a Gaussian random variable in Eq.~(\ref{eq:ad}).}
	\label{fig:amp_dist}	
\end{figure}

For the single-QD system $\mathcal A(1)$ Fig.~\ref{fig:amp_dist}(a) shows two examples, $\psi_{503}$ and $\psi_{1356}$, which are fairly well described by Eq.~(\ref{eq:ad}), with small deviations in the center and in the left and right wings. However, there are also many examples in this system that do not follow Eq.~(\ref{eq:ad}) (not shown), for instance the bouncing-ball state $\psi_{429}$ and the checkerboard-like states $\psi_{866},\psi_{1197}$; cf. Fig.~\ref{fig:unbound_1}. It is well known that such regular states are not described by a Gaussian random variable~\cite{MK88,Shigehara94}.

For the 4-QD system $\mathcal B$ we find that nearly all among twenty randomly chosen sample states between $\psi_{492},\ldots, \psi_{1581}$ show a remarkably good agreement with Eq.~(\ref{eq:ad}). Two examples are shown in Fig.~\ref{fig:amp_dist}(b).

Hence, as in the previous section, we find that the QD-WL model shows features known from nonintegrable and chaotic systems. Again, the system with more than one QD appears to be more \quoting{chaotic}, in the sense that its WL states behave more like Gaussian random variables. 
\begin{figure*}[tb]
\centering\includegraphics[width=0.8\linewidth]{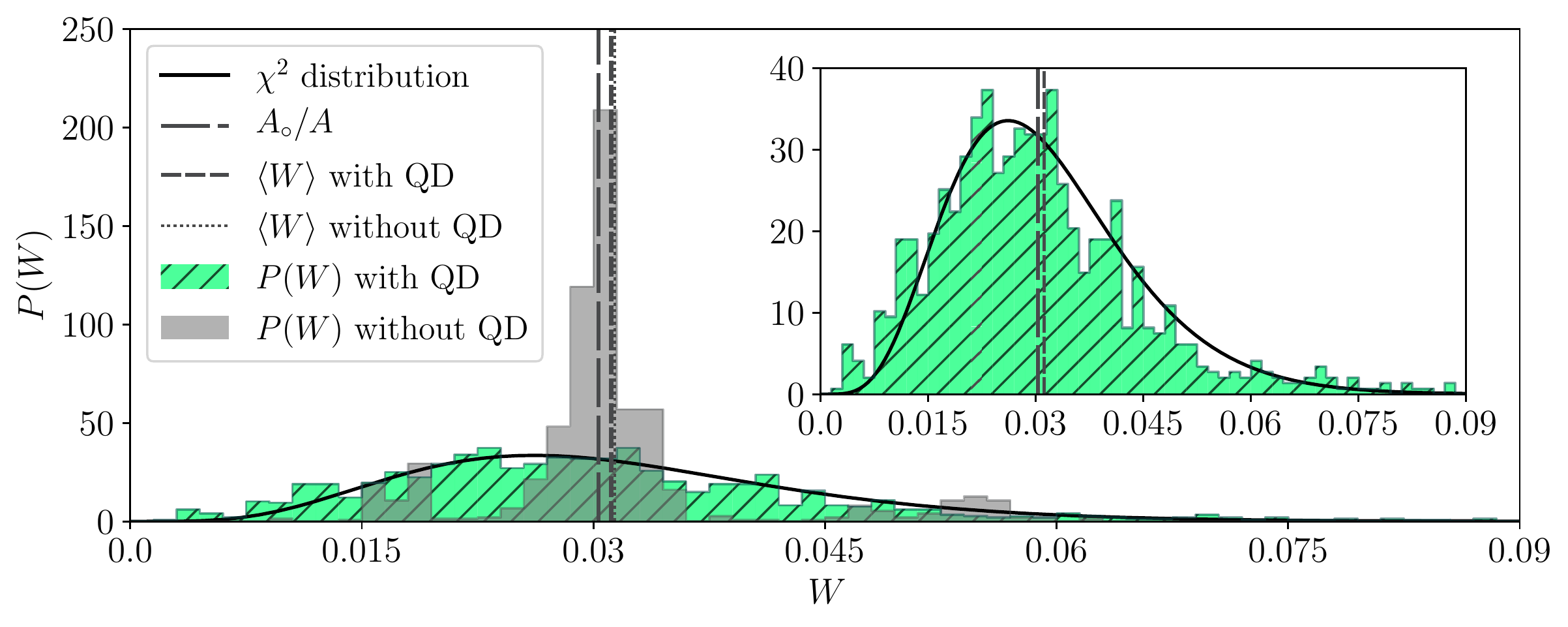}
	\caption{Probability distribution $P(W)$ of the localization $W$ [Eq.~(\ref{eq:loc})] of the WL states $i=4,\ldots,1000$ of the single-QD system $\mathcal A(1)$ with and without QD. Both distributions are normalized such that $\int_{0}^{W_{\text{max}}}dW P(W) = 1$ with $W_{\text{max}}=0.09$. Furthermore $\AQDs/\AWL \approx 0.03$ and $\mean{W}\approx 0.031$ for both distributions. The solid curve is the $\chi^2$ distribution~(\ref{eq:chi}) with $\nu=12$. The inset shows $P(W)$ for the case with QD on a smaller ordinate range.}
	\label{fig:loc_stat_1}	
\end{figure*}
\section{Localization}
\label{sec:loc_dist}
In this section we take up a question raised in Sec.~\ref{sec:morph} about the WL states' tendency to avoid or seek the region of the QD(s). For this purpose, we define the localization of the $i$-th eigenstate as
\begin{align}
W_i=\sum_{k=1}^{N_{\text{dot}}} \iint \limits_{B_k}^{} dy\, dx\ \vert\psi_i(x,y)\vert^2 \ ,
\label{eq:loc}
\end{align}
where, again, $B_k$ is the region where the confinement potential~$V_k$ of the $k$-th QD is nonzero. These double integrals are summed over all QDs and, like Eq.~(\ref{eq:doubleintegral}), are evaluated by using \textit{scipy.integrate.nquad} with upper bound on the number of subintervals being set to 100. 
For comparison we consider the ratio of the area of QDs ($\AQDs$) and the area of the WL ($A$) which can be written as
\begin{align}
\frac{\AQDs}{\AWL} = \frac{\pi }{4L_xL_y}\sum_{k=1}^{N_{\text{dot}}}b_k^2 \ .
\label{eq:arearatio}
\end{align}

For system $\mathcal A(1)$ we compute the localization $W_i$ for the WL states with $i=4,\ldots,1000$. We consider here a smaller amount of states than shown in Fig.~\ref{fig:unbound_1} due to the high sensitivity of the $W_i$ on the convergence of the series expansion [Eq.~(\ref{eq:expansion})], which here again contains $3000$ terms. For comparison, we calculate $W_i$ also for system $\mathcal A(1)$ under withdrawal of the QD having the plane wave-like energy eigenstates in Eq.~(\ref{eq:planewaves}). The resulting probability distributions $P(W)$ are depicted in Fig.~\ref{fig:loc_stat_1}. In the system without QD the distribution essentially shows one sharp peak at $\mean{W} \approx 0.03 \approx \AQDs/\AWL $ [see Eq.~(\ref{eq:arearatio})] quickly declining to zero and two flattened side lobes to its left and right. Inserting a QD considerably broadens the distribution $P(W)$ without changing the mean $\mean{W}$ which is therefore again very close to the area ratio in Eq.~(\ref{eq:arearatio}). 
That means a QD does not change the average localization but it does increase the fluctuations resulting in more WL states that tend to avoid the QD region but also more WL states that tend to seek it.

To explain the behavior of $P(W)$ in the QD-WL model we first note that if we would introduce a uniform absorption within the QD region, the resulting decay rate $\gamma_i$ of the $i$-th eigenstate would be in first-order perturbation theory for weak coupling proportional to the localization $W_i$. Hence, $\gamma_i/\mean{\gamma} = W_i/\mean{W}$. It has been shown that for weakly open systems (see, e.g.~\cite{GST18}) the lifetime statistics is given by a $\chi^2$ distribution 
\begin{align}
\chi^2_\nu(z) = \frac{(\nu/2)^{\nu/2}}{\Gamma\left(\nu/2\right)} z^{\nu/2-1} e^{-\nu z/2} \ ,
\label{eq:chi}
\end{align}
with normalized decay rate $z = \gamma_i/\mean{\gamma}$ and number of decay channels $\nu$. Equation~(\ref{eq:chi}) is the many-channel generalization of the Porter-Thomas distribution~\cite{PT56}. The $\chi^2$ distribution applies to integrable and chaotic systems provided that the coupling to the environment is weak and that the decay channels can be treated as independent Gaussian variables, see~\cite{GDM97} and references therein.
We therefore assume $P(W) \approx \chi^2_\nu( W/\mean{W})$ and use the number of decay channels $\nu$ as a fitting parameter. An upper bound for $\nu$ can be crudely estimated in the following way. First, we assume that the size of a decay channel in 2D real space is on average $(\lambda/2)^2$ with the wavelength $\lambda$ corresponding to the considered energy. Second, we count how many of such areas fit into the area~$\AQDs$ of the QD region. With the largest (unfolded) energy $\emax$ used in the statistics and Eq.~(\ref{eq:arearatio}) we find
\begin{align}
\numax = \emax \frac{4}{\pi}\frac{\AQDs}{A} \ .
\label{eq:numax}
\end{align}
It can be seen in Fig.~\ref{fig:loc_stat_1} that the numerically obtained $P(W)$ of $\mathcal A(1)$ is rather well fitted by the $\chi^2$ distribution for $\nu = 12$. This is consistent with the upper bound $\numax \approx 39$ using $\emax = 1000$.

The system $\mathcal A(1)$ without QD, see Fig.~\ref{fig:loc_stat_1}, cannot be well fitted by a $\chi^2$ distribution. We explain this by the fact that the decay channels here are not independent as the amplitudes of a regular state within a small QD region are strongly correlated. In particular, it means that a regular state cannot avoid to overlap with the QD region which implies that $P(W)$ is significantly reduced for small $W$.

\begin{figure}[tb]
\centering\includegraphics[width=1.\linewidth]{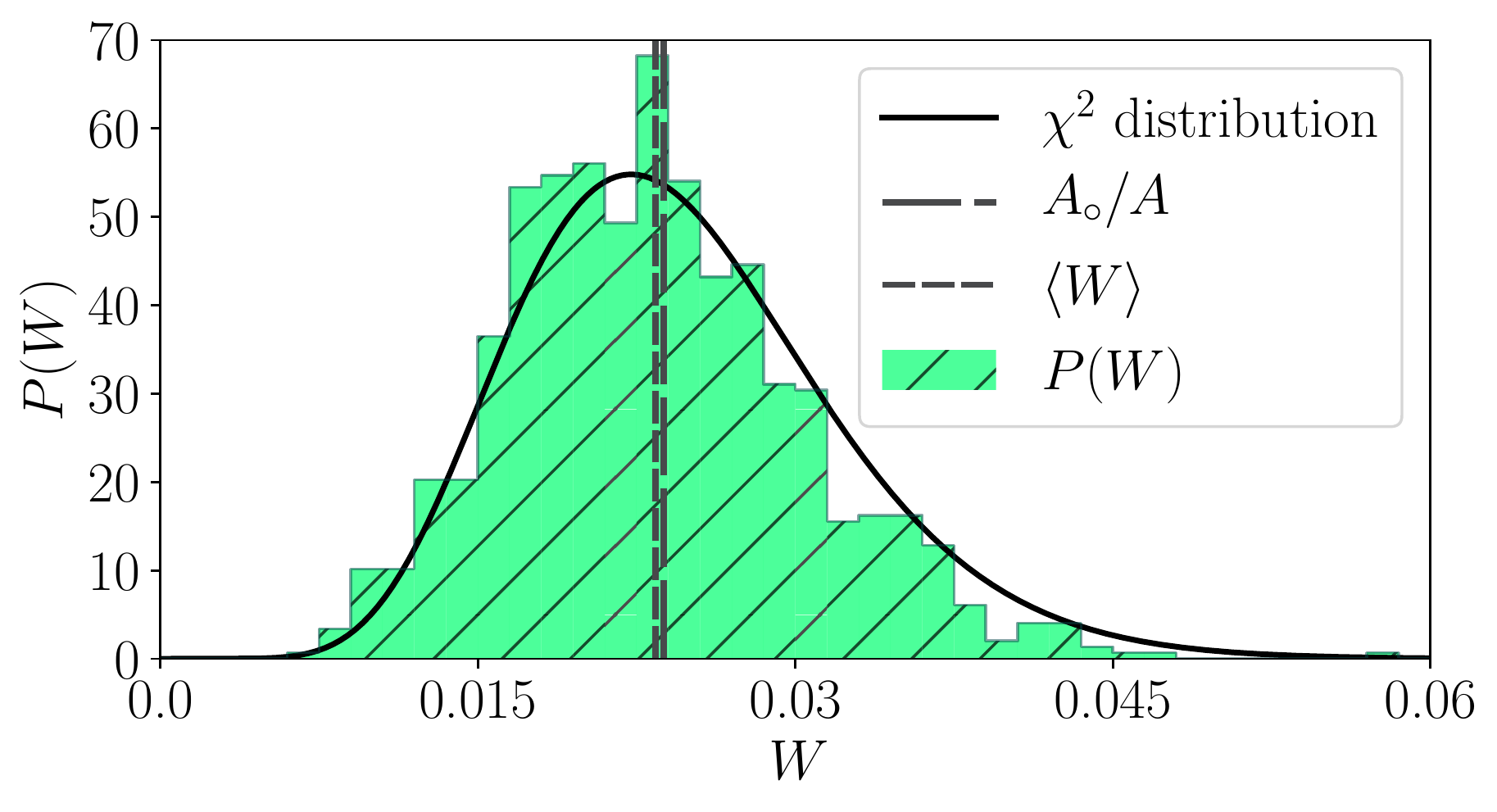}
	\caption{Probability distribution $P(W)$ of the localization $W$ of the WL states $i=13,\ldots,1000$ of 4-QD system $\mathcal B$. The distribution is normalized to unity with $W_{\text{max}}=0.06$ and compared to the $\chi^2$ distribution~(\ref{eq:chi}) with $\nu = 21$. Here, $\AQDs/\AWL \approx 0.0238$ and $\mean{W}\approx 0.0234$.}
	\label{fig:loc_stat_4QP}	
\end{figure}

Figure~\ref{fig:loc_stat_4QP} shows $P(W)$ for the 4-QD system $\mathcal B$. Here, the agreement with the $\chi^2$ distribution is even slightly better than for system $\mathcal A(1)$. The fitted value of $\nu = 21$ is consistent with the upper bound $\numax \approx 30$, using $\emax = 1000$ in Eq.~(\ref{eq:numax}). 

\begin{figure}[tb]
\centering\includegraphics[width=1.\linewidth]{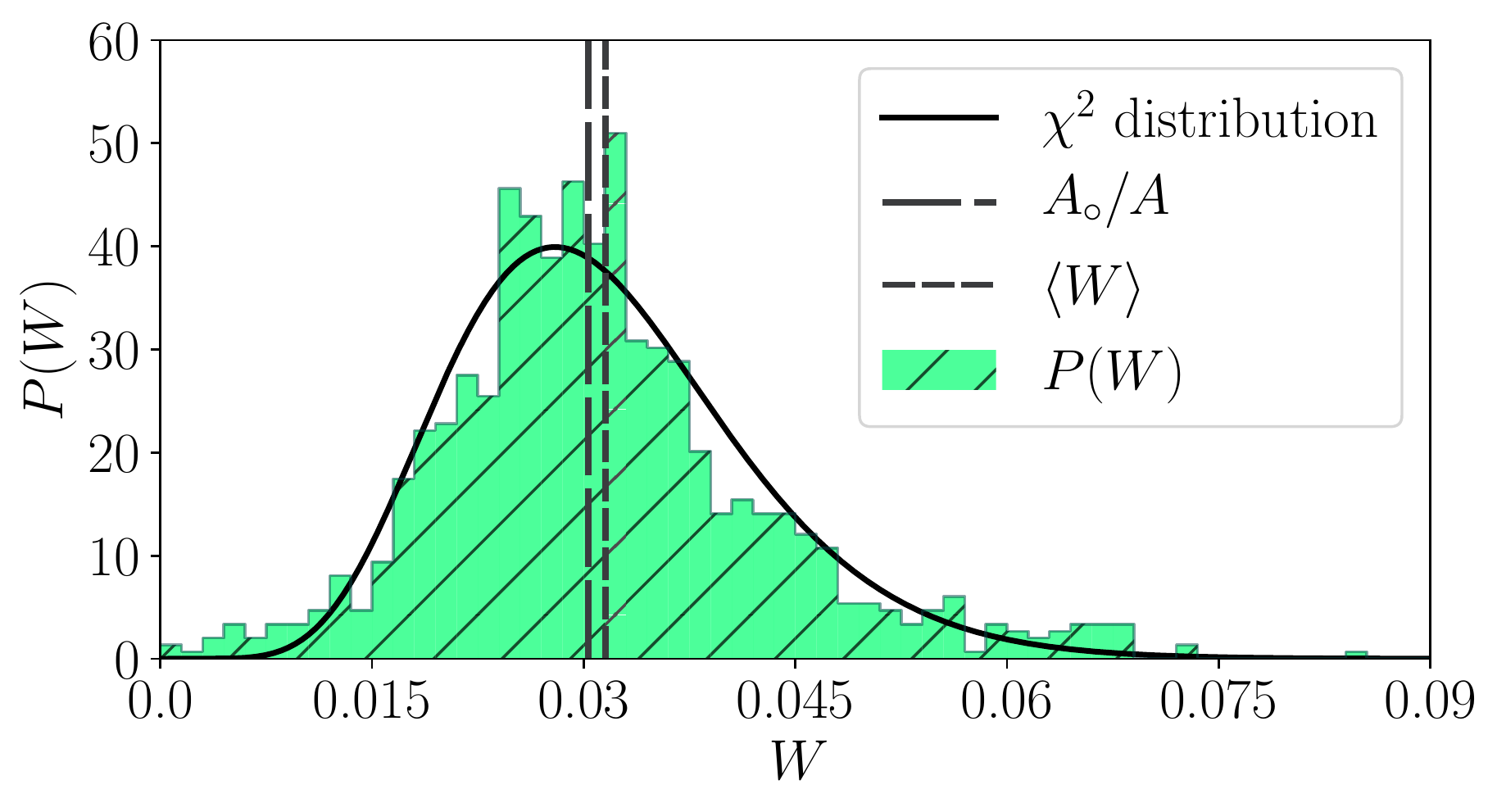}
 	\caption{Probability distribution $P(W)$ of the localization $W$ of the WL states $i=4,\ldots,1000$ of system $\mathcal A(1)$ for the same QD position, but for a different localization region, namely the region $B_1$ in Eq.~(\ref{eq:loc}) shifted to $(a_1,e_1) = (3.18,13.24)$. Like in Figs.~\ref{fig:loc_stat_1} and~\ref{fig:loc_stat_4QP} the distribution is normalized to unity and the $\chi^2$ distribution~(\ref{eq:chi}) with $\nu=18$ is shown as well. Here, $\mean{W}\approx 0.0315$.}
 	\label{fig:loc_stat_2}	
\end{figure}

Based on the so far observed statistical properties of the WL states it is natural to ask whether the localization in the QD(s) is different from the localization properties outside the QD(s). To answer this question we choose now $B_1$ in Eq.~(\ref{eq:loc}) for the single-QD system $\mathcal A(1)$ to be a spatially shifted region of the same shape and size, still in $\Omega$, but non-overlapping with the QD. 
Figure~\ref{fig:loc_stat_2} shows the resulting localization probability distribution with the shifted region centered at $(a_1,e_1) = (3.18,13.24)$. One can see that the general shape of $P(W)$ is as before with, however, an increased $\nu = 18$. Similar observations we have made for regions centered at $(a_1,e_1) = (7.52,11.74)$, $(2.64,2.57)$, and $(6.98,2.36)$. We therefore conclude that the statistical properties of the WL states are in first approximation independent of the position in the WL which is a feature shared with eigenstates in chaotic systems.

\section{Orthogonalized plane waves}
\label{sec:ortho}
\begin{figure}[tb]
\centering\includegraphics[width=0.9\linewidth]{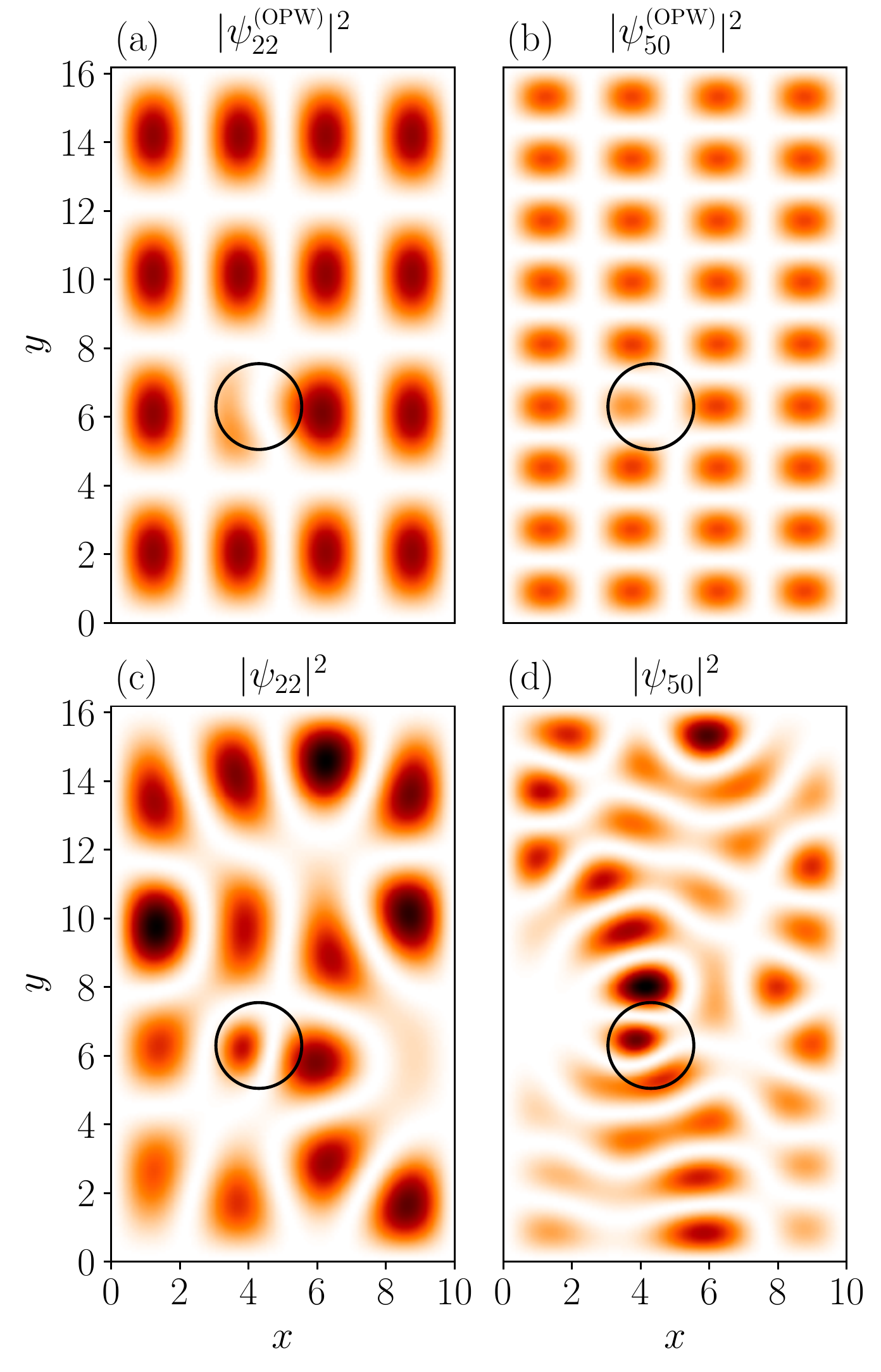}
	\caption{Comparison of OPWs [(a)-(b)] with corresponding, roughly regular-looking WL states [(c)-(d)], from series expansion~(\ref{eq:expansion}), containing $3000$ terms each, of system $\mathcal C$. The colormaps of (a) and (b) are adapted to their counterparts in (c) and (d), respectively. $x$ and $y$ are measured in nm.}
	\label{fig:schneider_1}	
\end{figure}
In this section we contrast the WL states with the OPWs. The OPW method tries to incorporate the impact of a QD on the WL states by orthogonalizing plane waves to the QD state(s). 
Following Ref.~\cite{Schneider01}, we consider the single-QD system $\mathcal C$ in Table~\ref{tab:para} which possess only one QD state due to a reduced parameter~$d_1$. This state is well approximated by the ground state of the isotropic 2D harmonic oscillator given by
\begin{align}
\langle x,y\vert\varphi_{\text{dot}}\rangle=C\, \exp{\left(-\frac{M\omega}{2\hbar} [(x-a_1)^2+(y-e_1)^2]\right)} \ ,
\label{eq:HO} 
\end{align}
with $C=\left(M\omega/\pi\hbar\right)^{1/2}$. In Eq.~(\ref{eq:HO}) the parameter $d_1$ is incorporated via the frequency $\omega=\sqrt{2d_1/M}$. 
The set of functions that need to be orthogonalized are those of the infinitely high potential well sorted in ascending order by their energy, with its ground state substituted by $\varphi_{\text{dot}}$:
\begin{align}
\{u_i\}_{i=1,\ldots,N}=\{u_1=\varphi_{\text{dot}},\ u_2=\varphi_2 ,\ldots,\ u_N=\varphi_N\} \ .
\label{on3}
\end{align}
It should be noted that in a set $\{u_i\}$ containing $u_1=\varphi_{1}$ the QD state $\varphi_{\text{dot}}$ could also constitute an additional state.

The approximated states are determined as
\begin{align}
\vert\psi_1^{(\text{OPW})}\rangle&=\frac{1}{N_1^{(\text{OPW})}}\vert u_1\rangle \ , \label{up68}\\
\vert{\psi}_i^{(\text{OPW})}\rangle&=\frac{1}{N_i^{(\text{OPW})}}\left(\vert u_i\rangle-\langle \psi_1^{(\text{OPW})}\vert u_i\rangle\vert \psi_1^{(\text{OPW})}\rangle\right)\ ,
\label{eq:ortho}
\end{align}
where $i=2,\ldots,N$ and $N_j^{(\text{OPW})}$ with $j = 1,\ldots,N$ is a normalization constant of the $j$-th state, such that the state is normalized to unity in $\Omega$. But unlike the exact eigenstates the OPWs $\psi_{i}^{(\mathrm{OPW})}$ are just orthogonal to the ground state $\psi_1^{(\mathrm{OPW})}$
\begin{align}
\langle \psi_1^{(\mathrm{OPW})}\vert\psi_i^{(\mathrm{OPW})}\rangle&=0 \ , \quad i\geq 2
\label{on6}
\end{align}
but not necessarily to each other.

Two examples of OPWs are shown in Figs.~\ref{fig:schneider_1}(a) and~(b). As expected from the above orthogonalization procedure, these states are plane waves but with a modification in the QD region. It is obvious that such OPWs cannot approximate the complex morphology of WL states in Figs.~\ref{fig:unbound_1} and~\ref{fig:states_2}. 
Figures~\ref{fig:schneider_1}(c) and (d) show two exact WL states with a roughly regular, checkerboard-like structure. In these specific examples, it seems that the OPWs in Figs.~\ref{fig:schneider_1}(a) and~(b) almost can approximate the WL states. However, the wave function inside the respective QD region, which is important for the carrier capture rate calculations~\cite{Schneider01}, is not correctly predicted.

There is another troublesome effect becoming noticeable for increasing energy. While in the exact treatment of the QD-WL model even for single-QD systems the QD does not lose its impact on the WL states even for rather high energies, in the OPW approach the impact quickly lessens until it finally ceases. This results from a decreasing overlap of $\langle \psi_1^{(\text{OPW})}\vert u_i\rangle$ in Eq.~(\ref{eq:ortho}) due to stronger spatial oscillations of the $u_{i}$ for higher energies. 

As the OPW method fails here to reproduce the complex pattern of WL states, we expect that this approach does not describe the carrier capture relaxation process into QDs better than the simpler approach of using just plane waves. This statement is consistent with the findings in Ref.~\cite{MKT06}.

\section{Conclusion}
\label{sec:conc}
In this paper we have studied the properties of wetting-layer states in a simple quantum-dot-wetting-layer model. 
The spectral analysis via the nearest-neighbor spacing distribution and the number variance showed that both the short- and long-range level correlations are intermediate between Poisson statistics and the Gaussian orthogonal ensemble of random-matrix theory, with a tendency towards the latter when the size of the wetting layer or the number of quantum dots is increased.
By dividing the spectra into distinct energy windows we demonstrated that the nearest-neighbor spacing distribution shows a slow transition towards Poisson statistics when the energy is increased.

Moreover, we demonstrated numerically that the spatial structure of the  wetting-layer states shows a complex morphology. Depending on the number of dots we observed distinct classes of wetting-layer states. In the simplest case of a rectangular-shaped wetting layer with one quantum dot the possible behavior ranges from checkerboard-like states up to states with unequally or more irregular pattern. 
By increasing the number of quantum dots we observed that checkerboard-like states disappear. We showed that the irregular looking states exhibit probability amplitude distributions very similar to that of a Gaussian random variable.

Furthermore, we investigated the localization behavior of wetting-layer states in the quantum-dot region(s). Due to the presence of a dot, the probability distribution of the localization is broadened without changing the mean value. Hence, the average localization does not change but there are more states that tend to avoid the quantum-dot region but also more states that tend to seek it. Studying the localization in some distance from the quantum dot revealed a similar behavior. 
Interestingly, the localization distribution in the presence of quantum dot(s) is well fitted by a $\chi^2$ distribution. 

The observed localization behavior can be important for capture time calculations, such as in Ref.~\cite{Schneider01}. In this context, we demonstrated that the orthogonalized plane waves are not capable of accurately approximating the wetting-layer states.

In the studied parameter regime we have not seen any signatures of Anderson localization~\cite{Anderson58} which could be observable in our model for a very large wetting-layer area with many randomly-placed quantum dots. 

In this work we considered quantum dots on a wetting layer of finite rectangular shape. Even this simple shape, which in the absence of any quantum dot would have plane wave-like energy eigenstates, give rise to wetting-layer states with a complex morphology. We showed that increasing the size of the wetting layer even increases the complexity of the wetting-layer states in terms of their level statistics. We expect that more complicated shapes of the wetting layer that might appear in experiments do not change the qualitative behavior as the energy eigenstates in the absence of quantum dots already have an irregular spatial structure, see, e.g.~\cite{Stoeckmann00}. Following the same line of reasoning, we also expect that deviations of the quantum-dot confinement potential from the truncated parabolic shape do not change the qualitative behavior of the wetting-layer states.

Based on the insights presented in this work, we surmise that an improved approach to approximate wetting-layer states may consist in using random superpositions of plane waves. The statistical properties of such states, see, e.g.~\cite{OGH87}, are closer to those of the wetting-layer states. We therefore expect that random superpositions of plane waves are more tailored to the needs of the actual wetting-layer states than (orthogonalized) plane waves.

\section*{Acknowledgments}
We would like to thank J. Kullig, B. Gulyak, and G. Kasner for helpful discussions.

\bibliographystyle{aipnum4-1}

\end{document}